# Quantum Spin-1/2 Rings Built from [2]Triangulene Molecular Units

*Can Li[1,#,*], Manish Kumar[3,#], Ying Wang[2,4#], Diego Manuel Soler Polo[3], Yi-Jun Wang[2], He Qi[4], Liang Liu[1,5,6], Xiaoxue Liu[1,5,6], Dandan Guan[1,5,6], Yaoyi Li[1,5,6], Hao Zheng[1,5,6], Canhua Liu[1,5,6], Jinfeng Jia[1,5,6], Pei-Nian Liu[2,*], Pavel Jelinek[3,*], Deng-Yuan Li[2,*], and Shiyong Wang[1,5,6,*]*

[1]State Key Laboratory of Micro-nano Engineering Science, Key Laboratory of Artificial Structures and Quantum Control (Ministry of Education), Tsung-Dao Lee Institute, School of Physics and Astronomy, Shanghai Jiao Tong University, Shanghai 200240, China

[2]Key Laboratory of Natural Medicines, School of Pharmacy, China Pharmaceutical University, Nanjing 211198, China

[3]Institute of Physics, Czech Academy of Sciences, Prague 16200, Czech Republic

[4]School of Chemistry and Molecular Engineering, East China University of Science & Technology, Shanghai, 200237, China

[5]Hefei National Laboratory, Hefei 230088, China

[6]Shanghai Research Center for Quantum Sciences, 99 Xiupu Road, Shanghai 201315, China

ABSTRACT

Quantum spin rings represent fundamental model systems that exhibit distinctive quantum phenomena—such as quantum critical behavior and quasiparticle excitations—arising from their periodic boundary conditions and enhanced quantum fluctuations. Here, we report the on-surface synthesis and atomic-scale characterization of antiferromagnetic S=1/2 quantum spin rings composed of pristine and unmodified [2]triangulene units on a Au(111) surface. Using stepwise on-surface synthesis followed by STM tip-induced dehydrogenation, we precisely constructed cyclic five- and six-membered spin rings and investigated their spin states via scanning probe microscopy and multireference calculations. Nc-AFM imaging reveals that the six-membered ring retains a planar geometry, whereas the five-membered ring exhibits pronounced structural distortion. The six-membered ring hosts a uniform excitation gap that can be accurately described by a Heisenberg spin model and multireference CASCI calculations. In contrast, the distorted five-membered ring displays spin ground states with asymmetric spatial distributions due to degeneracy lifting induced by structural distortion. Our findings establish a versatile molecular platform for exploring correlated magnetism and quantum spin phenomena in cyclic organic magnetic architectures with disorder.

INTRODUCTION

Open-shell nanographenes provide an ideal platform for investigating exotic magnetic phenomena driven by topological frustration, sublattice imbalance, and electron correlations.[1-26] In nanographene architectures, these interactions generate unpaired π-electrons whose highly delocalized spins lead to unconventional magnetic behavior.[27] When organized into one-dimensional chains, these spin units exhibit rich quantum collective phenomena described by Heisenberg or Haldane models. Specifically, finite $S=1$ spin chains can undergo quantum topological phase transitions and host fractionalized end states,[8,14] while $S=1/2$ spin chains with open boundary conditions demonstrate characteristic spinon quasiparticle excitations.[1-3,6,7]

The topological transition from spin chains to spin ring geometries introduces periodic boundary conditions, fundamentally altering the system's excitation spectrum and collective spin behavior.[11,16,18,28,29] In $S=1/2$ spin rings, substrate interactions can be modulated by peripheral functional groups,[16] while in $S=1$ systems, next-nearest-neighbor couplings significantly influence spin excitation energies.[18,29] These characteristics make quantum spin rings invaluable for probing low-dimensional quantum magnetism, quantum entanglement, and frustration in strongly correlated systems. However, a critical yet unexplored aspect in molecular quantum spin ring systems is how adsorption-induced dihedral deformations affect spin properties—a key structural factor that remains unexamined. Although on-surface synthesis now allows atomic-precision fabrication of spin-based nanostructures, experimental challenges persist, particularly in locally characterizing intrinsic $S=1/2$ spin rings and their subtle structure-spin relationships.

Here, we achieve the precise construction of cyclic five- and six-membered quantum spin rings with unsubstituted [2]triangulene on Au(111) by combining on-surface synthesis and scanning tunneling microscopy (STM) tip-induced dehydrogenation and characterize their spin properties by scanning tunneling spectroscopy (STS). Each [2]triangulene unit hosts a delocalized spin-1/2 moment (Figure 1a) originating from sublattice imbalance,[12,13,20,27,30,31] enabling us to systematically trace the evolution of spin excitations from quantum spin chains to quantum spin rings. Our findings reveal that the transition from spin chains to spin rings, governed by periodic boundary conditions, leads to significant enhancement of both ground-state degeneracy and spin excitation gaps. Furthermore, molecular geometry—particularly dihedral distortions between neighboring [2]triangulene units—introduces disordered spin exchange interactions,[10,17,32-34] resulting in ground-state degeneracy lifting and spatial localization of spin density. Combined with multireference complete active space configuration interaction (CASCI) calculations (see SOM in Ref. 35), we reveal a fundamental distinction between the two architectures: the planar hexamer maintains symmetric spin excitations characteristic of uniform coupling, whereas the pentamer undergoes a geometry-driven lifting of its ground-state degeneracy. This symmetry breaking manifests as strongly asymmetric spatial distributions in both Kondo resonances and spin-flip excitations, providing directly visualizing the impact of molecular geometry on spin correlations and frustration. These findings establish engineered molecular spin rings as a versatile platform for constructing low-dimensional quantum magnets and exerting precise control over their spin excitation.

# RESULTS AND DISCUSSION

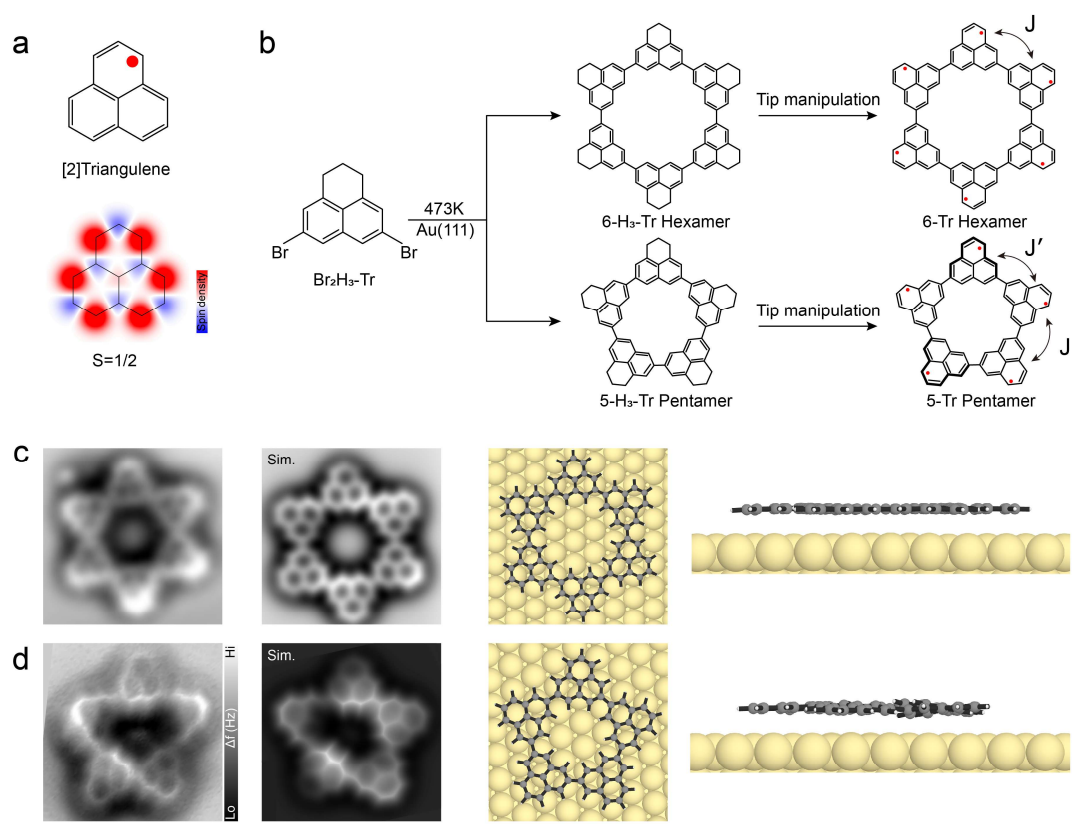

**Figure 1.** On-surface synthesis and characterization of quantum spin rings. (a) Chemical structure and calculated spin density distribution of an open-shell [2]triangulene molecule obtained using the mean-field Hubbard model. (b) Schematic representation of the on-surface synthetic route leading from closed-shell to open-shell molecular rings with tip-dehydrogenation. (c-d) Experimental bond-resolved nc-AFM images and corresponding simulated AFM images[36] based on DFT-optimized adsorption structures of the planar 6-Tr hexamer and the buckled 5-Tr pentamer on Au(111), shown in both top and side views.

**On-surface synthesis and topography properties of molecular spin rings.** Owing to the exceptional reactivity of open-shell [2]triangulene, the direct synthesis of [2]triangulene-based spin architectures on surfaces is often hampered by uncontrolled side reactions.[3] To circumvent this issue, we design and synthesize a closed-shell molecular precursor, 5,8-dibromo-2,3-dihydro-1*H*-phenalene (Br$_2$H$_3$-Tr) by solution methods (see the detailed synthesis and structural characterization in the Figure S1), where the [2]triangulene core passivated by three hydrogen atoms ($H_3$-Tr) is expected to have suitable reactivity on surfaces (Figure 1b). The precursor $Br_2H_3$-Tr were deposited onto a clean Au(111) substrate under ultrahigh vacuum conditions and annealed to 473 K to activate the debromination C-C coupling reaction. The results shown the formation of well-defined linear polymer chains and cyclic oligomers with five and six $H_3$-Tr units, such as pentamer and hexamer (labeled as 5-$H_3$-Tr pentamer and 6-$H_3$-Tr hexamer). In this work, we focus on the cyclic pentamer and hexamer.

Next, we introduced unpaired π-electron spins into the molecular cyclic systems through site-selective dehydrogenation induced by an STM tip.[4] This was achieved by applying a ~2.4 V pulse precisely over the $sp^3$-hybridized carbon site of each $H_3$-Tr unit, which removed the hydrogen atoms and generated a [2]triangulene with a local S = 1/2. Figure S2 displays the as-synthesized closed-shell 5-$H_3$-Tr pentamer and 6-$H_3$-Tr hexamer before dehydrogenation, where the bright protrusions correspond to the $sp^3$-hybridized carbon atoms of $H_3$-Tr units. After STM tip-induced dehydrogenation, these $sp^3$ carbon sites revert to $sp^2$ hybridization, leading to planarized structures (Figure 1c,d). The formation of open-shell [2]triangulene within both cyclic spin systems(labeled as 5-Tr pentamer or 6-Tr hexamer) is unambiguously confirmed by bond-resolved noncontact atomic force microscopy (nc-AFM).

Notably, the 6-Tr hexamer adopt a planar geometry, whereas the 5-Tr pentamer exhibit pronounced buckling due to the dihedral angles between neighboring units, as corroborated by simulated nc-AFM images[36] obtained for fully optimized molecule on Au(111) surface using density functional theory (DFT) calculations[37] shown in Figure 1c,d. This structural distinction may result in disordered spin-exchange coupling in the pentamers,[33,34] in contrast to the uniform coupling in the hexamers.

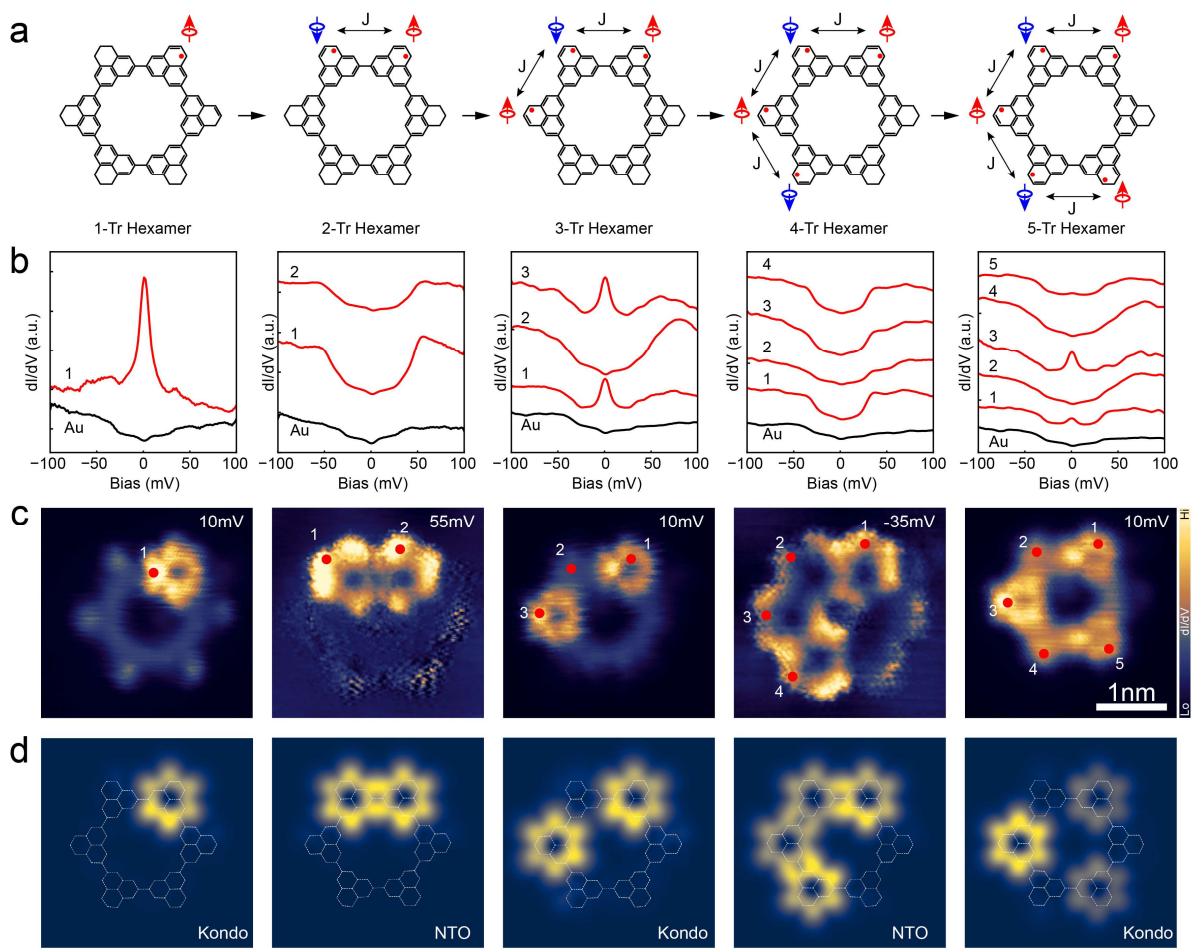

**Figure 2.** Spin excitations in a cyclic hexamer. (a) Schematic illustrating the sequential incorporation of one to five spins into the molecular ring. (b) Site-specific dI/dV spectra measured at the positions marked in (c). (c) Experimental spatial distributions of the Kondo resonance for the 1-, 3-, and 5-spin hexamers, and inelastic spin excitations for the 2- and 4-spin hexamers. (d) Simulated dI/dV maps showing the spatial distribution of Kondo orbitals for the 1-, 3-, and 5-spin hexamers, and natural transition orbitals for the 2- and 4-spin hexamers.

**Tracing the evolution of spin excitations in molecular rings.** We have systematically traced the evolution of spin states and magnetic excitations in synthetic molecular rings, from the initial creation of local moments to the emergence of collective quantum behaviors (Figures 2a and S10). Starting from a closed-shell hexamer structure, we sequentially introduced localized S=1/2 spins via STM tip-induced dehydrogenation, as illustrated in Figure 2a. This stepwise approach allows us to observe the transition from isolated local moments to antiferromagnetically coupled chains, and finally to a complete quantum spin ring.

Introducing a single spin into the 6-$H_3$-Tr forming a 1-Tr hexamer. STS characterization on this unit reveals a pronounced zero-bias Kondo resonance (Figure 2b, left panel), confirming a well-defined local magnetic moment S=1/2. The spatial mapping of the Kondo signal shows that it is delocalized throughout the entire [2]triangulene unit, consistent with the simulated dI/dV map of calculated Kondo orbitals[38] (Figure 2d), while being absent on neighboring closed-shell sites.

Adding a second spin at an adjacent site quenches the Kondo resonance and gives rise to a step-like feature in the dI/dV spectrum at 44 meV (Figure 2b). This feature is characteristic of a spin-flip excitation from a singlet ground state to a triplet excited state. The energy of this step corresponds directly to the antiferromagnetic exchange coupling strength, $J \approx 44$ meV. This assignment is further confirmed by the second-derivative inelastic tunneling spectroscopy($d^2I/dV^2$) in the Supporting Information (Figure S3). To further visualize the spatial characteristics of this coupled spin system, we performed dI/dV mapping at the inelastic excitation energy. The spatial distribution of this inelastic signal is delocalized over both spin sites. As shown in Figure 2c, the resulting experimental spatial distribution of the spin excitation closely resembles simulated dI/dV maps of the natural transition orbitals (NTO)[39,40] calculated from multireference CASCI calculation corresponding to the spin-excitation from open-shell singlet to triplet excitation which are delocalized on both the [2]triangulene unit (Figures S13 and S14).

The incorporation of a third spin into the system yields a doublet ground state. This is evidenced by the reappearance of zero-bias Kondo peaks on the first and third sites, despite all three sites hosting an unpaired electron (Figure 2b, c). Generating a fourth spin, the system returns to a singlet ground state, and the Kondo resonance is absent. Only spin-flip excitations are observed. The spatial distribution of the inelastic signal is non-uniform, showing higher intensity on the terminal (1st and 4th) sites compared to the central pair (2nd and 3rd) (Figure 2c). The 5-Tr system mirrors the behavior of the 3-Tr chain: it possesses a doublet ground state and exhibits a Kondo resonance localized only on the odd-numbered sites. The spatial map of its spin excitation also shows a stronger signal on the even-numbered sites. The spatial distributions are corroborated by multireference CASCI calculations (Figures 2d and S11 to S22).

The resulting spin excitations across these 2-Tr hexamer to 5-Tr hexamer can also be described by the antiferromagnetic Heisenberg model for quantum spin-1/2 chains. A clear parity effect emerges between chains with odd and even numbers of spins. In odd-length chains, the system retains a net spin-1/2 moment, resulting in a doublet ground state. This leads to a distinct zero-bias Kondo peak near the Fermi level, which is absent in even-numbered chains—where only spin-flip excitations are observed. To capture the spin-excitation behaviors, we modeled the interacting π-electron spins using a Heisenberg Hamiltonian with a uniform coupling strength of $J \approx 44$ meV:

$$H = J \sum_{<i,i+1>} S_i \cdot S_{i+1}$$

The calculated excitation energies presented in Figure S6 agree well with the experimental data. For both 2-Tr and 3-Tr hexamer, the first excited state lies at approximately 44 meV, matching the antiferromagnetic coupling strength J extracted from the measurements. In the 4-Tr and 5-Tr hexamer, the excitation energy decreases to about 30 meV, consistent with the trend observed in the dI/dV spectra. These results confirm that the tip-generated π-electron spins in the cyclic hexamer realize an antiferromagnetic Heisenberg quantum spin-1/2 system with up to five interacting spins.

**Effect of cyclic geometry on spin excitations in hexamer spin ring.** Incorporation of the sixth spin closes the molecular structure into a quantum spin ring, thereby affecting a topological transition from an open chain to a system governed by periodic boundary conditions. This topological closure not only substantially modifies the low-energy spin excitations but also dramatically alters the quasiparticle behavior in the entire excitation spectrum.[41]

In this cyclic 6-Tr hexamer system, the perfectly planar molecular geometry ensures a nearly uniform exchange coupling J between all neighboring sites. The ground state is confirmed as an open-shell singlet, as demonstrated by the lack of a Kondo resonance and the clear observation of spin-flip excitations in the dI/dV spectra (Figure 3c). Remarkably, the measured excitation energy exceeds that of the corresponding 6-spin open chain, a finding well reproduced by theoretical calculations using uniform Heisenberg spin chain and ring models (Figure S9).

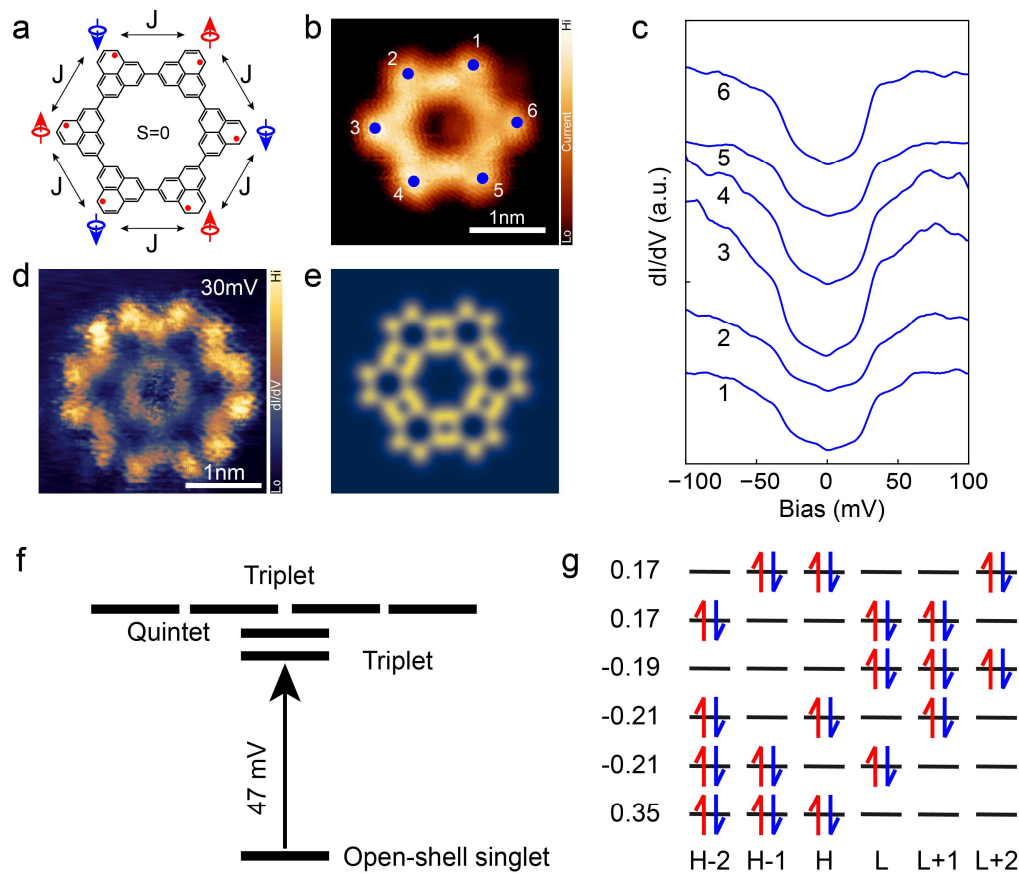

**Figure 3.** Spin excitations in a cyclic hexamer. (a) Schematic illustration of the cyclic hexamer. (b) STM topographic images of the corresponding open-shell molecular structures. (c) Site-specific dI/dV spectra measured at the positions marked in (b). (d) Experimental dI/dV maps recorded at spin-excitation step. (e) Theoretically simulated dI/dV map of the natural transition orbitals corresponding to the spectra shown in (d). (f) Calculated spectra of planar hexamer showing open-shell singlet ground state with a few excited states. (g) multireference wavefunction of the open-shell singlet ground state of the hexamer.

This enhancement is a direct consequence of the ring geometry. Our multireference CASCI calculations for the planar hexamer (Figure 3f) reveal an open-shell singlet ground state, with the first excited state being a triplet at 47 meV. This excitation gap is followed by a manifold of nearly degenerate quintet and higher triplet states. This theoretical prediction aligns perfectly with the experimental spectra, where the first sharp step is followed by a broad, V-like feature, indicative of multiple excitations within a narrow energy window (Figure 3c). The calculated energy spectrum for a uniform Heisenberg ring model confirms that periodic boundary conditions elevate the spin excitation energies and introduce additional degeneracies compared to the open-chain analogue (Figure S9).

Inelastic spin excitation dI/dV mapping at 30mV (Figure 3d) reveals a homogeneous distribution across all six [2]triangulene units. This uniform spatial profile is accurately captured by the simulated dI/dV maps[42] using calculated NTOs for the dominant open-shell singlet to triplet excitation (Figure 3e). Furthermore, the open-shell singlet ground state (Figure 3g) exhibits full multireference wavefunction with a combination of many Slater determinants, consistent with a highly symmetric and strongly correlated quantum spin ring

**Spin states of cyclic five-membered spin rings.** W In contrast to the symmetric hexamer, the cyclic pentamer presents an example of geometric spin frustration. Its odd-numbered ring geometry inherently prevents perfect antiferromagnetic order, leading to degenerated ground states (Figure 4c). However, our experiments reveal that this ideal picture is altered by structural distortions, which lift degeneracies and introduce non-uniform exchange interaction between adjacent spins.

We first consider the theoretical model of a perfectly planar pentamer. Multireference CASCI calculations for this idealized structure reveal the spin frustration in the ground and excited state (Figure 4c). The ground state is a doubly degenerate doublet ($\Psi_0^D, \Psi_1^D$), indicating that the

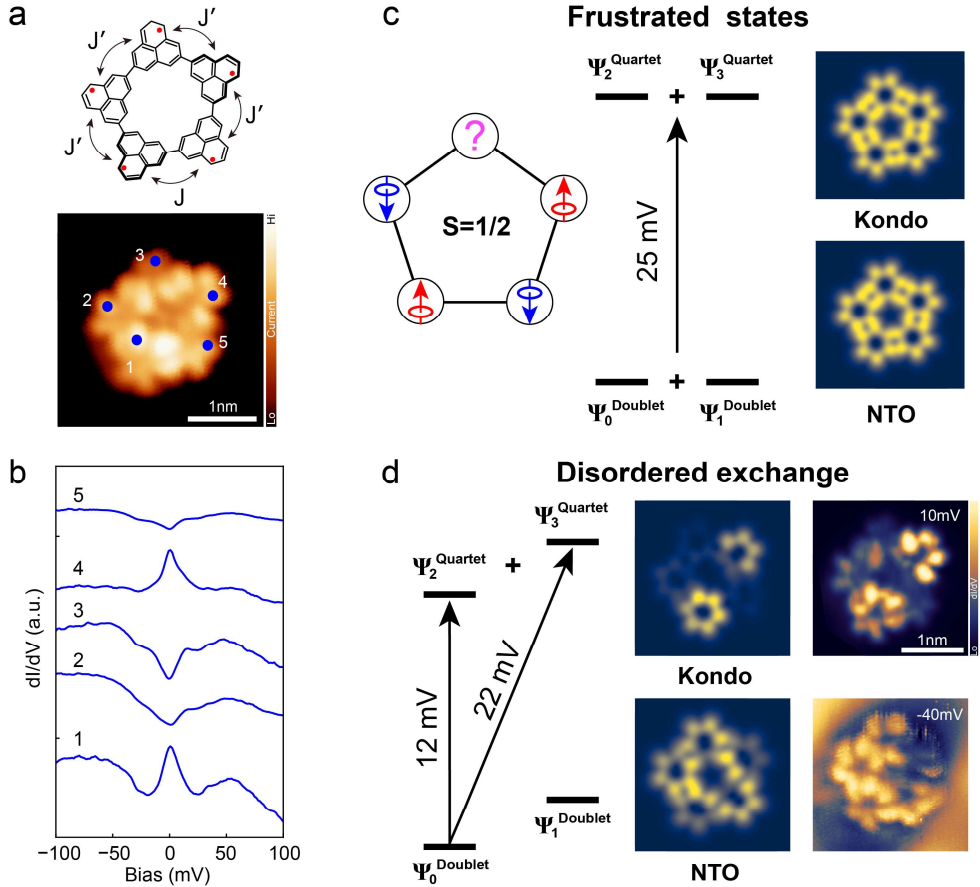

**Figure 4.** Spin excitations in a cyclic pentamer. (a) Schematic illustration of the cyclic pentamer with STM topographic images of the corresponding open-shell molecular structures. (b) Site-specific dI/dV spectra measured at the positions marked in STM topographic image in (a). (c) Energy spectra of fully planer pentamer from CASCI calculations showing doubly degenerated doublet ground state followed by doubly degenerated quartet state with simulated dI/dV maps of Kondo and NTO. (d) Energy spectra obtained from CASCI calculation for the pentamer using relaxed geometry on the gold surface showing the lifting of degeneracy of doublet and quartet states with the corresponding experimental dI/dV maps of Kondo and spin-excitation and also the simulated dI/dV maps of Kondo and NTO.

five spins cannot form a unique, stable antiferromagnetic arrangement. The first excited state is also doubly degenerate, comprising a quartet state ($\Psi_0^Q, \Psi_1^Q$) found 25 meV above the ground state. This electronic degeneracy has direct spectroscopic consequences. Calculations of the Kondo orbitals,[38] which account for the degenerate ground states, predict a perfectly symmetric Kondo resonance on every [2]triangulene site. Similarly, the simulated spatial dI/dV map[42] of the spin-excitation signal, derived from the NTOs connecting the degenerate doublet and quartet states, where molecule can go from the first doublet to first quartet ($\Psi_0^D \rightarrow \Psi_1^Q$), first doublet to second quartet ($\Psi_0^D \rightarrow \Psi_2^Q$), second doublet to first quartet ($\Psi_1^D \rightarrow \Psi_0^Q$), and second doublet to second quartet ($\Psi_1^D \rightarrow \Psi_1^Q$) is also uniform (Figure 4c). In this ideal, high-symmetry case, the spin frustration manifests as electronic degeneracy but preserves the spatial symmetry of all observable signals

Due to steric hindrance between the hydrogen atoms of adjacent units, our observed five-membered spin ring is distorted with significant dihedral angles (5-20º) between adjacent units (Figure 1d). This structural buckling breaks the molecular point group symmetry and modulates an effective exchange interaction between adjacent triangulene units.[33,34] This symmetry breaking is directly observed in the spectroscopy. Site-specific dI/dV spectra (Figure 4b) show a pronounced spatial heterogeneity, a clear zero-bias Kondo resonance appears only on sites 1 and 4, while the other three sites exhibit only spin-flip features. Furthermore, the intensity of these spin-excitation steps varies significantly from site to site.

This behavior arises from a disorder in the exchange coupling J induced by the varying dihedral angles.[33,34] Our DFT calculations on [2]triangulene dimers confirm that the exchange coupling J decreases significantly as the dihedral angle increases (Figure S7). Consequently, in the adsorbed pentamer, each pair of neighboring spins has a distinct coupling strength $J_{i,i+1}$, effectively creating a Hamiltonian with spatially modulated exchange.[32]

The CASCI calculations for the pentamer geometry-relaxed on surface also confirm this picture. The structural distortion lifts the degeneracies of the ideal case: the ground state is now a single non-degenerate doublet, and the excited quartet states split (Figure 4d). The calculated Kondo orbitals[38] for this non-degenerate ground state are localized predominantly on two sites, in agreement with the experimental observation of Kondo resonances on sites 1 and 4. Similarly, the simulated NTO map for the spin-excitation now shows a highly asymmetric spatial distribution, mirroring the experimental dI/dV map (Figure 4d). Thus, the pentamer demonstrates a compelling interplay between geometry and magnetism. While its odd-membered ring geometry inherently suggests spin frustration, the actual magnetic ground state is dictated by molecular conformation. The steric hindrance-induced buckling breaks the symmetry, lifts the degeneracy, and transforms the system from a uniformly frustrated ring into a spin system with non-uniform exchange interactions, localizing spin density and creating an asymmetric excitation landscape.

CONCLUSION

In summary, we have demonstrated the bottom-up assembly and atomic-scale characterization of quantum spin rings constructed from [2]triangulene molecular building blocks. Combining scanning probe microscopy/spectroscopy with multireference CASCI calculations, we unveil the evolution of spin states from individual local moments to collective quantum phases. Our findings reveal a fundamental transition in spin excitation behavior upon ring closure—from open Heisenberg spin chains to rings with periodic boundary conditions. The planar cyclic hexamer exhibits uniform antiferromagnetic coupling, giving rise to enhanced excitation energies and characteristic degeneracies of a symmetric quantum spin ring. In contrast, the non-planar pentamer displays geometrically modulated exchange interactions, where dihedral distortions lift the ground-state degeneracy and induce spin localization. This work establishes a versatile molecular platform for exploring quantum magnetism in ring geometries and opens new avenues for the bottom-up design of correlated spin systems with tailored quantum states.

METHODS SECTION

**Sample Preparation and STM/nc-AFM Measurement.** A commercial low-temperature Unisoku Joule-Thomson scanning probe microscope (4.2 K) operated at ultrahigh vacuum ($3 \times 10^{-10}$ mbar) was used for all sample preparation and characterization. The Au(111) single crystals were cleaned using repeated cycles of $Ar^+$ sputtering and subsequent annealing to 900 K to obtain atomically flat terraces. The cleanliness of the crystals was checked by scanning with the STM before molecular deposition. Precursor 5,8-dibromo-2,3-dihydro-1*H*-phenalene was thermally deposited on the clean Au(111) surface at room temperature and annealed to 473K for 10 minutes. Precursor 5,8-dibromo-2,3-dihydro-1*H*-phenalene was evaporated on the surface from a quartz crucible and the sublimation temperature was approximately 300 K. Carbon monoxide molecules were dosed onto the cold sample around 9 K ($1\times10^{-8}$ mbar, 1 minute). A lock-in amplifier (571 Hz, 30 mV modulation) has been used to obtain dI/dV spectra and mappings. The STM, STS, and nc-AFM measurements were taken at 4.2 K, and the images were processed using the WSxM software. To achieve ultra-high spatial resolution, CO molecule was picked up from the Au(111) surface to the apex of a tungsten tip. A quartz tuning fork with a resonant frequency of 26.4 kHz has been used for nc-AFM measurements. The sensor was operated in frequency modulation mode with a constant oscillation amplitude of 100 pm. AFM measurements were performed in constant-height mode with Bias = 3 mV.

**CASCI calculations.** The molecular geometries were optimized on a three-layer Au(111) slab, with the bottom gold layer kept fixed, using density functional theory (DFT) as implemented in the FHI-aims software package[37]. The PBE exchange-correlation functional[43] was employed, and van der Waals interactions were accounted for using the Tkatchenko–Scheffler method[44]. Given the open-shell and multi-radical character of the molecules under study, the complete active space configuration interaction (CASCI) method was used to accurately describe their wavefunctions and electronic energies.[35] The one- and two-electron integrals, constructed in the basis of molecular orbitals around the Fermi energy, were derived using the ORCA quantum chemistry package,[45] with orbitals obtained from restricted open-shell Kohn-Sham (ROKS) calculations. For the CASCI calculations, the geometry of the molecule was taken from the structure optimized on the Au(111) surface to include surface-induced effects on molecular geometry. An active space of 11 electrons in 11 orbitals [CAS(11,11)] was used for molecules with an odd number of electrons, while 12 electrons in 12 orbitals [CAS(12,12)] were used for those with an even number of electrons. To determine the number of unpaired electrons in the molecule, we calculated the occupation of natural orbitals by diagonalizing the one-particle density matrix constructed from the ground-state many-body CASCI wavefunction.

**Kondo Orbitals (KO).** Kondo orbitals are calculated by diagonalizing the Hamiltonian derived from the multi-channel Anderson model, which considers the many-body multiplet structure of molecules obtained from the CASCI calculation for the neutral ground state and virtual charge states as described in the Ref. 38.

**Natural Transition Orbitals (NTOs).** We calculated NTOs[39] for the spin-flip process between the ground state and the first excited state, in order to capture the spatial variation of the dI/dV maps corresponding to IETS spin excitation maps.

Simulated nc-AFM and dI/dV images were obtained using the probe particle code,[36,42] which employed optimized geometries derived from total energy DFT calculations.

ASSOCIATED CONTENT

**Supporting Information**.

The Supporting Information is available free of charge at http://pubs.acs.org.

Detailed descriptions of precursor synthesis, and expanded additional STM/nc-AFM and theoretical calculation results (PDF)

AUTHOR INFORMATION

**Corresponding Authors**

\***Can Li** – *State Key Laboratory of Micro-nano Engineering Science, Key Laboratory of Artificial Structures and Quantum Control (Ministry of Education), Tsung-Dao Lee Institute,*

*School of Physics and Astronomy, Shanghai Jiao Tong University, Shanghai 200240, China;*
Email: lic_18@sjtu.edu.cn

**\*Pei-Nian Liu** − *Key Laboratory of Natural Medicines, School of Pharmacy, China Pharmaceutical University, Nanjing 211198, China;* Email: liupn@cpu.edu.cn

**\*Pavel Jelinek** – *Institute of Physics, Czech Academy of Sciences, Prague 16200, Czech Republic;* Email: jelinekp@fzu.cz.

**\*Deng-Yuan Li** − *State Key Laboratory of Natural Medicines, School of Pharmacy, China Pharmaceutical University, Nanjing, 211198, China;* Email: dengyuanli@cpu.edu.cn

**\*Shiyong Wang** – *State Key Laboratory of Micro-nano Engineering Science, Key Laboratory of Artificial Structures and Quantum Control (Ministry of Education), Tsung-Dao Lee Institute, School of Physics and Astronomy, Shanghai Jiao Tong University, Shanghai 200240, China; Hefei National Laboratory, Hefei 230088, China; Shanghai Research Center for Quantum Sciences, 99 Xiupu Road, Shanghai 201315, China;* Email: shiyong.wang@sjtu.edu.cn.

**Authors**

**Manish Kumar** – *Institute of Physics, Czech Academy of Sciences, Prague 16200, Czech Republic*

**Ying Wang** – *State Key Laboratory of Natural Medicines, School of Pharmacy, China Pharmaceutical University, Nanjing, 211198, China; School of Chemistry and Molecular Engineering, East China University of Science & Technology, Shanghai, 200237, China*

**Diego Manuel Soler Polo** – *Institute of Physics, Czech Academy of Sciences, Prague 16200, Czech Republic*

**Yi-Jun Wang** – *State Key Laboratory of Natural Medicines, School of Pharmacy, China Pharmaceutical University, Nanjing, 211198, China*

**He Qi** – *School of Chemistry and Molecular Engineering, East China University of Science & Technology, Shanghai, 200237, China*

**Liang Liu** – *State Key Laboratory of Micro-nano Engineering Science, Key Laboratory of Artificial Structures and Quantum Control (Ministry of Education), Tsung-Dao Lee Institute, School of Physics and Astronomy, Shanghai Jiao Tong University, Shanghai 200240, China; Hefei National Laboratory, Hefei 230088, China; Shanghai Research Center for Quantum Sciences, 99 Xiupu Road, Shanghai 201315, Chinaa*

**Xiaoxue Liu** – *State Key Laboratory of Micro-nano Engineering Science, Key Laboratory of Artificial Structures and Quantum Control (Ministry of Education), Tsung-Dao Lee Institute, School of Physics and Astronomy, Shanghai Jiao Tong University, Shanghai 200240, China; Hefei National Laboratory, Hefei 230088, China; Shanghai Research Center for Quantum Sciences, 99 Xiupu Road, Shanghai 201315, China*

**Dandan Guan** – *State Key Laboratory of Micro-nano Engineering Science, Key Laboratory of Artificial Structures and Quantum Control (Ministry of Education), Tsung-Dao Lee Institute, School of Physics and Astronomy, Shanghai Jiao Tong University, Shanghai 200240, China; Hefei National Laboratory, Hefei 230088, China; Shanghai Research Center for Quantum Sciences, 99 Xiupu Road, Shanghai 201315, China*

**Yaoyi Li** – *State Key Laboratory of Micro-nano Engineering Science, Key Laboratory of Artificial Structures and Quantum Control (Ministry of Education), Tsung-Dao Lee Institute, School of Physics and Astronomy, Shanghai Jiao Tong University, Shanghai 200240, China;*

*Hefei National Laboratory, Hefei 230088, China; Shanghai Research Center for Quantum Sciences, 99 Xiupu Road, Shanghai 201315, China*

**Hao Zheng –** *State Key Laboratory of Micro-nano Engineering Science, Key Laboratory of Artificial Structures and Quantum Control (Ministry of Education), Tsung-Dao Lee Institute, School of Physics and Astronomy, Shanghai Jiao Tong University, Shanghai 200240, China; Hefei National Laboratory, Hefei 230088, China; Shanghai Research Center for Quantum Sciences, 99 Xiupu Road, Shanghai 201315, China*

**Canhua Liu –** *State Key Laboratory of Micro-nano Engineering Science, Key Laboratory of Artificial Structures and Quantum Control (Ministry of Education), Tsung-Dao Lee Institute, School of Physics and Astronomy, Shanghai Jiao Tong University, Shanghai 200240, China; Hefei National Laboratory, Hefei 230088, China; Shanghai Research Center for Quantum Sciences, 99 Xiupu Road, Shanghai 201315, China*

**Author Contributions**

D.-Y.L. and S.W. conceived and supervised the experiments; C.L. and S.W. performed the STM/STS and nc-AFM measurements under the supervision of J.F.J.; D.-Y.L. designed the molecular precursor; Y.W., Y.-J. W., and H. Q. synthesized the precursor molecules with the supervision of P.-N.L.; C.L. and S.W. carried out the Heisenberg calculations; P. J., M. K. and D.S.-P. carried out DFT calculations. C.L., D.-Y.L. and S.W. wrote the manuscript; All authors discussed the results and commented on the manuscript at all stages. All authors have given approval to the final version of the manuscript. #These authors contributed equally.

**Notes**

The authors declare no competing financial interests.

ACKNOWLEDGMENT

We thank the NSFC (Grants No.22325203, No.92577203, No.92365302, No.92265105, No.92065201, No.92580139, No.12074247, No.12174252, No.22272050, No.12304230, No.22161160319), the National Key R\&D Program of China (Grants No.2024YFF0727103, Grants No.2020YFA0309000), the Strategic Priority Research Program of Chinese Academy of Sciences (Grant No.XDB28000000) and the Science and Technology Commission of Shanghai Municipality (Grants No.2019SHZDZX01, No.20QA1405100, No.24LZ1401000, No.22QA1403000), Cultivation Project of Shanghai Research Center for Quantum Sciences (Grant No.LZPY2024-04), Innovation program for Quantum Science and Technology (Grant No.2021ZD0302500), and Shanghai Jiao Tong University 2030 Initiative for financial support. P.J., D.S.P. and M.K acknowledge financial support from the CzechNanoLab Research Infrastructure supported by MEYS CR (LM2023051) and the Czech Science Fundation project no. 25-17866X.

REFERENCES

(1) Su, X.; Ding, Z.; Hong, Y.; Ke, N.; Yan, K.; Li, C.; Jiang, Y.-F.; Yu, P. Fabrication of spin-1/2 Heisenberg antiferromagnetic chains via combined on-surface synthesis and reduction for spinon detection. *Nat. Synth* **2025**, 4, 694–701.

(2) Fu, X.; Huang, L.; Liu, K.; Henriques, J. C. G.; Gao, Y.; Han, X.; Chen, H.; Wang, Y.; Palma, C.-A.; Cheng, Z.; Lin, X.; Du, S.; Ma, J.; Fernández-Rossier, J.; Feng, X.; Gao, H.-

J. Building spin-1/2 antiferromagnetic Heisenberg chains with diazananographenes. *Nat. Synth* **2025**, 4, 684–693.

(3) Yuan, Z.; Zhang, X.-Y.; Jiang, Y.; Qian, X.; Wang, Y.; Liu, Y.; Liu, L.; Liu, X.; Guan, D.; Li, Y.; Zheng, H.; Liu, C.; Jia, J.; Qin, M.; Liu, P.-N.; Li, D.-Y.; Wang, S. Fractional spinon quasiparticles in open-shell triangulene spin-1/2 chains. *J. Am. Chem. Soc.* **2025**, 147, 5004–5013.

(4) Li, C.; Wang, Y.; Jiang, Y.; Liu, Y.; Wu, Y.; Han, Y.-J.; Chen, Z.-Q.; Liu, L.; Liu, X.; Guan, D.; Li, Y.; Zheng, H.; Liu, C.; Liu, P.-N.; Jia, J.; Li, D.-Y.; Wang, S. Programmable higher-order topological phases in open-shell metal–organic frameworks. *J. Am. Chem. Soc.* **2025**, 147, 39662–39670.

(5) Zheng, Y.; Li, C.; Xu, C.; Beyer, D.; Yue, X.; Zhao, Y.; Wang, G.; Guan, D.; Li, Y.; Zheng, H.; Liu, C.; Liu, J.; Wang, X.; Luo, W.; Feng, X.; Wang, S.; Jia, J. Designer spin order in diradical nanographenes. *Nat. Commun.* **2020**, 11, 6076.

(6) Zhao, C.; Catarina, G.; Zhang, J.-J.; Henriques, J. C. G.; Yang, L.; Ma, J.; Feng, X.; Gröning, O.; Ruffieux, P.; Fernández-Rossier, J.; Fasel, R. Tunable topological phases in nanographene-based spin-1/2 alternating-exchange Heisenberg chains. *Nat. Nanotechnol.* **2024**, 19, 1789–1795.

(7) Zhao, C.; Yang, L.; Henriques, J. C. G.; Ferri-Cortés, M.; Catarina, G.; Pignedoli, C. A.; Ma, J.; Feng, X.; Ruffieux, P.; Fernández-Rossier, J.; Fasel, R. Spin excitations in nanographene-based antiferromagnetic spin-1/2 Heisenberg chains. *Nat. Mater.* **2025**, 24, 722–727.

(8) Zhao, Y.; Jiang, K.; Li, C.; Liu, Y.; Zhu, G.; Pizzochero, M.; Kaxiras, E.; Guan, D.; Li, Y.; Zheng, H.; Liu, C.; Jia, J.; Qin, M.; Zhuang, X.; Wang, S. Quantum nanomagnets in on-surface metal-free porphyrin chains. *Nat. Chem.* **2023**, 15, 53–60.

(9) Zhao, Y.; Jiang, K.; Liu, P.-Y.; Li, J.; Li, R.; Li, X.; Fang, X.; Zhao, A.; Zhu, Y.; Xu, H.; Chen, T.; Wang, D.; Zhuang, X.; Hou, S.; Wu, K.; Gao, S.; Sun, Q.-F.; Zhang, Y.; Wang, Y. Construction of Kondo chains by engineering porphyrin -radicals on Au(111). *J. Am. Chem. Soc.* **2025**, 147, 38100–38109.

(10) Yu, H.; Jing, Y.; Heine, T. Physics and chemistry of two-dimensional triangulene-based lattices. *Acc. Chem. Res.* **2025**, 58, 61–72.

(11) Su, J.; Fan, W.; Mutombo, P.; Peng, X.; Song, S.; Ondráček, M.; Golub, P.; Brabec, J.; Veis, L.; Telychko, M.; Jelínek, P.; Wu, J.; Lu, J. On-surface synthesis and characterization of [7]triangulene quantum ring. *Nano Lett.* **2021**, 21, 861–867.

(12) Mishra, S.; Beyer, D.; Eimre, K.; Liu, J.; Berger, R.; Gröning, O.; Pignedoli, C. A.; Müllen, K.; Fasel, R.; Feng, X.; Ruffieux, P. Synthesis and characterization of -extended triangulene. *J. Am. Chem. Soc.* **2019**, 141, 10621–10625.

(13) Mishra, S.; Xu, K.; Eimre, K.; Komber, H.; Ma, J.; Pignedoli, C. A.; Fasel, R.; Feng, X.; Ruffieux, P. Synthesis and characterization of [7]triangulene. *Nanoscale* **2021**, 13, 1624–1628.

(14) Mishra, S.; Catarina, G.; Wu, F.; Ortiz, R.; Jacob, D.; Eimre, K.; Ma, J.; Pignedoli, C. A.; Feng, X.; Ruffieux, P.; Fernández-Rossier, J.; Fasel, R. Observation of fractional edge excitations in nanographene spin chains. *Nature* **2021**, 598, 287–292.

(15) Liu, Y.; Weigold, S.; Yan, L.; Wei, Z.; Hanne, M.; Tverskoy, O.; You, S.; Xie, M.; Zhang, Y.; Chen, Q.; Rominger, F.; Bunz, U. H. F.; Freudenberg, J.; Du, S.; Müllen, K.; Chi, L. Steering magnetic coupling in diradical nonbenzenoid nanographenes. *J. Am. Chem. Soc.* **2025**, 147, 23103–23112.

(16) Li, D.; Cao, N.; Metzelaars, M.; Silveira, O. J.; Jestilä, J.; Fumega, A.; Nishiuchi, T.; Lado, J.; Foster, A. S.; Kubo, T.; Kawai, S. Frustration-induced many-body degeneracy in spin-1/2 molecular quantum rings. *J. Am. Chem. Soc.* **2025**, 147, 26208–26217.

(17) Krane, N.; Turco, E.; Bernhardt, A.; Jacob, D.; Gandus, G.; Passerone, D.; Luisier, M.; Juríček, M.; Fasel, R.; Fernández-Rossier, J.; Ruffieux, P. Exchange interactions and intermolecular hybridization in a spin-1/2 nanographene dimer. *Nano Lett.* **2023**, 23, 9353–9359.

(18) Hieulle, J.; Castro, S.; Friedrich, N.; Vegliante, A.; Lara, F. R.; Sanz, S.; Rey, D.; Corso, M.; Frederiksen, T.; Pascual, J. I.; Peña, D. On-surface synthesis and collective spin excitations of a triangulenebased nanostar. *Angew. Chem. Int. Ed.* **2021**, 60, 25224–25229.

(19) Daugherty, M. C.; Jacobse, P. H.; Jiang, J.; Jornet-Somoza, J.; Dorit, R.; Wang, Z.; Lu, J.; McCurdy, R.; Tang, W.; Rubio, A.; Louie, S. G.; Crommie, M. F.; Fischer, F. R. Regioselective on-surface synthesis of [3]triangulene graphene nanoribbons. *J. Am. Chem. Soc.* **2024**, 146, 15879–15886.

(20) Su, J.; Telychko, M.; Hu, P.; Macam, G.; Mutombo, P.; Zhang, H.; Bao, Y.; Cheng, F.; Huang, Z.-Q.; Qiu, Z.; Tan, S. J. R.; Lin, H.; Jelínek, P.; Chuang, F.-C.; Wu, J.; Lu, J. Atomically precise bottom-up synthesis of -extended [5]triangulene. *Sci. Adv.* **2019**, 5, eaav7717.

(21) Mishra, S.; Beyer, D.; Eimre, K.; Ortiz, R.; Fernández-Rossier, J.; Berger, R.; Gröning, O.; Pignedoli, C. A.; Fasel, R.; Feng, X.; Ruffieux, P. Collective all-carbon magnetism in triangulene dimers. *Angew. Chem. Int. Ed.* **2020**, 132, 12139–12145.

(22) Li, C.; Liu, Y.; Liu, Y.; Xue, F.-H.; Guan, D.; Li, Y.; Zheng, H.; Liu, C.; Jia, J.; Liu, P.-N.; Li, D.-Y.; Wang, S. Topological defects induced high-spin quartet state in truxene-based molecular graphenoids. *CCS Chem.* **2023**, 5, 695–703.

(23) Song, S.; Su, J.; Telychko, M.; Li, J.; Li, G.; Li, Y.; Su, C.; Wu, J.; Lu, J. On-surface synthesis of graphene nanostructures with -magnetism. *Chem. Soc. Rev.* **2021**, 50, 3238–3262.

(24) Song, S.; Pinar Solé, A.; Matěj, A.; Li, G.; Stetsovych, O.; Soler, D.; Yang, H.; Telychko, M.; Li, J.; Kumar, M.; Chen, Q.; Edalatmanesh, S.; Brabec, J.; Veis, L.; Wu, J.; Jelinek, P.; Lu, J. Highly entangled polyradical nanographene with coexisting strong correlation and topological frustration. *Nat. Chem.* **2024**, 16, 938–944.

(25) Li, E.; Kumar, M.; Peng, X.; Shen, T.; Soler-Polo, D.; Wang, Y.; Teng, Y.; Zhang, H.; Song, S.; Wu, J., et al. Designer polyradical nanographenes with strong spin entanglement and perturbation resilience via Clar's goblet extension. *arXiv:*2506.05181 **2025**.

(26) Zuzak, R.; Kumar, M.; Stoica, O.; Soler-Polo, D.; Brabec, J.; Pernal, K.; Veis, L.; Blieck, R.; Echavarren, A. M.; Jelinek, P., et al. On-Surface Synthesis and Determination of the Open-Shell Singlet Ground State of Tridecacene. *Angew. Chem. Int. Ed.* **2024**, 63, e202317091.

(27) Lieb, E. H. Two theorems on the Hubbard model. *Phys. Rev. Lett.* **1989**, 62, 1201.

(28) Villalobos, F. et al. Globally aromatic odd-electron π-magnetic macrocycle. *Chem* **2025**, 11, 102316.

(29) Zhu, X.; Jiang, Y.; Wang, Z.; Huang, Y.; Luo, Z.; Yan, K.; Wang, S.; Yu, P. Collective Magnetism of Spin Coronoid via On-Surface Synthesis. *J. Am. Chem. Soc.* **2025**, 147, 10045–10051.

(30) Pavliček, N.; Mistry, A.; Majzik, Z.; Moll, N.; Meyer, G.; Fox, D. J.; Gross, L. Synthesis and characterization of triangulene. *Nat. Nanotechnol.* **2017**, 12, 308–311.

(31) Wang, T.; Berdonces-Layunta, A.; Friedrich, N.; Vilas-Varela, M.; Calupitan, J. P.; Pascual, J. I.; Peña, D.; Casanova, D.; Corso, M.; Oteyza, D. G. D. Aza-triangulene: On-surface synthesis and electronic and magnetic properties. *J. Am. Chem. Soc.* **2022**, 144, 4522–4529.

(32) Catarina, G.; Turco, E.; Krane, N.; Bommert, M.; Ortega-Guerrero, A.; Gröning, O.; Ruffieux, P.; Fasel, R.; Pignedoli, C. A. Conformational tuning of magnetic interactions in coupled nanographenes. *Nano Lett.* **2024**, 24, 12536–12544.

(33) Yu, H.; Heine, T. Magnetic coupling control in triangulene dimers. *J. Am. Chem. Soc.* **2023**, 145, 19303–19311.

(34) Pérez-Elvira, E.; Lozano, M.; Huang, Q.; Ma, J.; Gallardo, A.; Barragán, A.; Lauwaet, K.; Gallego, J. M.; Miranda, R.; Jelínek, P.; Écija, D.; Soler-Polo, D.; Feng, X.; Urgel, J. I. Reactivity and Magnetic Coupling of Triangulene Dimers Linked via para-Biphenyl Units. *Angew. Chem. Int. Ed.* **2025**, 137, e202501874.

(35) Biswas, K. et al. Steering Large Magnetic Exchange Coupling in Nanographenes near the Closed-Shell to Open-Shell Transition. *J. Am. Chem. Soc.* **2023**, 145, 2968–2974.

(36) Hapala, P.; Kichin, G.; Wagner, C.; Tautz, F. S.; Temirov, R.; Jelínek, P. Mechanism of high-resolution STM/AFM imaging with functionalized tips. *Phys. Rev. B* **2014**, 90, 085421.

(37) Blum, V.; Gehrke, R.; Hanke, F.; Havu, P.; Havu, V.; Ren, X.; Reuter, K.; Scheffler, M. Ab initio molecular simulations with numeric atom-centered orbitals. *Comput. Phys. Commun.* **2009**, 180, 2175–2196.

(38) Calvo-Fernández, A.; Kumar, M.; Soler-Polo, D.; Eiguren, A.; Blanco-Rey, M.; Jelínek, P. Theoretical model for multiorbital Kondo screening in strongly correlated molecules with several unpaired electrons. *Phys. Rev. B* **2024**, 110, 165113.

(39) Martin, R. L. Natural transition orbitals. *J. Chem. Phys.* **2003**, 118, 4775–4777.

(40) Kumar, M.; Soler-Polo, D.; Lozano, M.; Monino, E.; Veis, L.; Jelinek, P. Multireference Theory of Scanning Tunneling Spectroscopy Beyond One-Electron Molecular Orbitals: Can We Image Molecular Orbitals? *J. Am. Chem. Soc.* **2025**, 147, 24993–25003.

(41) Henriques, J. C. G.; Zhao, C.; Catarina, G.; Ruffieux, P.; Fasel, R.; Fernández-Rossier, J. Determining energy dispersion of spin excitations with scanning tunneling spectroscopy. *Phys. Rev. Lett.* **2025**, 135, 096703.

(42) Krejčí, O.; Hapala, P.; Ondráček, M.; Jelínek, P. Principles and simulations of high-resolution STM imaging with a flexible tip apex. *Phys. Rev. B* **2017**, 95, 045407.

(43) Perdew, J. P.; Burke, K.; Ernzerhof, M. Generalized gradient approximation made simple. *Phys. Rev. Lett.* **1996**, 77, 3865.

(44) Tkatchenko, A.; DiStasio, R. A.; Car, R.; Scheffler, M. Accurate and Efficient Method for Many-Body van der Waals Interactions. *Phys. Rev. Lett.* **2012**, 108, 236402.

(45) Neese, F. The ORCA program system. *Wiley Interdisciplinary Reviews: Computational Molecular Science* **2012**, 2, 73–78.

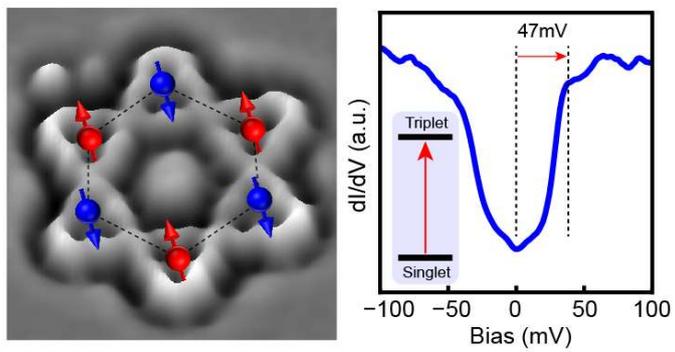